\documentclass{llncs}
    
\let\llncssubparagraph\subparagraph
\let\subparagraph\paragraph
\usepackage[compact]{titlesec}
\let\subparagraph\llncssubparagraph

\usepackage{times}
\usepackage{epsfig}
\usepackage{graphicx}
\usepackage{amsmath}
\usepackage{amssymb,pifont}

\usepackage{tabularx}
\usepackage{diagbox}
\usepackage{multirow}

\usepackage{booktabs}
\usepackage{upgreek}
\usepackage{makecell}
\usepackage{authblk}

\usepackage[table]{xcolor}
\usepackage{hyperref}
\hypersetup{colorlinks=false,citecolor=green, final}

\definecolor{grey}{rgb}{0.33, 0.33, 0.33}
\definecolor{red}{rgb}{1,0,0}
\definecolor{blue}{rgb}{0,0,1}

\newcolumntype{C}{>{\centering\arraybackslash}X}

\begin{document}
\frontmatter          
\pagestyle{headings}  
%
%
\mainmatter              
\title{Patch-based Non-Local Bayesian Networks for Blind Confocal Microscopy Denoising}
\authorrunning{Izadi et al.} 
%

%

\author{Saeed Izadi, and Ghassan Hamarneh}
\institute{
School of Computing Science, Simon Fraser University, Canada \\
\email{\{saeedi, hamarneh\}@sfu.ca}\\
}

\maketitle              

\newcommand{\cmark}{\ding{51}} 
\newcommand{\xmark}{\ding{53}}

\newcommand\tblResults{
\begin{table*}[t!]
\vspace{-3mm}
\caption{Quantitative results of NLBNN-P and NLBNN-S against baseline methods in term of PSNR$\pm$SEM and SSIM averaged over noisy images of the held-out FOVs}
\label{tab:results}
\begin{center}
\begin{tabular}{|l|c|c|c|c|c|c|c|c|}
\Xhline{1pt}
\rowcolor[rgb]{ .851,  .851,  .851}

Dataset
&
\multicolumn{2}{c|}{BPAE}
&
\multicolumn{2}{c|}{Mouse Brain}
&
\multicolumn{2}{c|}{Zebrafish}
\\
\hline
\rowcolor[rgb]{ .851,  .851,  .851}

Metric&PSNR$\uparrow$&SSIM$\uparrow$
&PSNR$\uparrow$&SSIM$\uparrow$
&PSNR$\uparrow$&SSIM$\uparrow$
\\
\hline
\hline

NLM~\cite{nlm2005}
&34.74 & 0.9108 & 36.31 & 0.9534 & 28.23 & 0.7895
\\
BM3D~\cite{bm3d2007}
&35.86 & 0.9338 & 37.95 & 0.9637 & 32.00 & 0.8854
\\
N2S~\cite{n2s2019}
& 36.01 & 0.9388 & 37.49 & .9574 & 32.14 & 0.8889
\\
PN2V~\cite{krull2019probabilistic}
& - & - & 38.24 & - & 32.45 & -
\\
N2N~\cite{n2n2018}
& 36.35 & 0.9441 & 38.19 & 0.9665 & 32.93 & 0.9076
\\
DnCNN~\cite{dncnn2017}
&36.12 & 0.9399 & 38.14 & 0.9686 & 32.29 & 0.9001
\\
\hline
\hline
NLBNN-S
&35.94 & 0.9388 & 37.74 & 0.9611 & 32.18 & 0.8986
\\
NLBNN-P
&36.02 & 0.9398 & 38.12 & 0.9631 & 32.48 & 0.9036

\\
\Xhline{1pt}
\end{tabular}
\vspace{-7mm}
\end{center}

\label{t2}
\end{table*}
}

\newcommand\tblNoiseLevel{
\begin{table}[t!]
\vspace{-3mm}
\caption{Quantitative results of WhiteNNer against baseline methods for different noise levels. 
\label{tab:noiselevel}
}
\begin{center}
\resizebox{\columnwidth}{!}{%
\begin{tabular}{|l||c|c|c|}
\Xhline{1pt}
\rowcolor[rgb]{ .851,  .851,  .851}
\multicolumn{1}{|c||}{Dataset}
&BPAE-B
&BPAE-G
&BPAE-R

\\
\rowcolor[rgb]{ .851,  .851,  .851}
\hline
\multicolumn{1}{|c||}{Metric}

&PSNR$\uparrow$
&PSNR$\uparrow$
&PSNR$\uparrow$
\\
\hline
\hline
NLBNN-L
&-&-&-
\\
NLBNN-NL
&-&-&-
\\
\hline
\end{tabular}
}
\vspace{-7mm}
\end{center}
\end{table}
}

\newcommand\figArch{
\begin{figure*}[t!]

\centering
  \includegraphics[width=\textwidth]{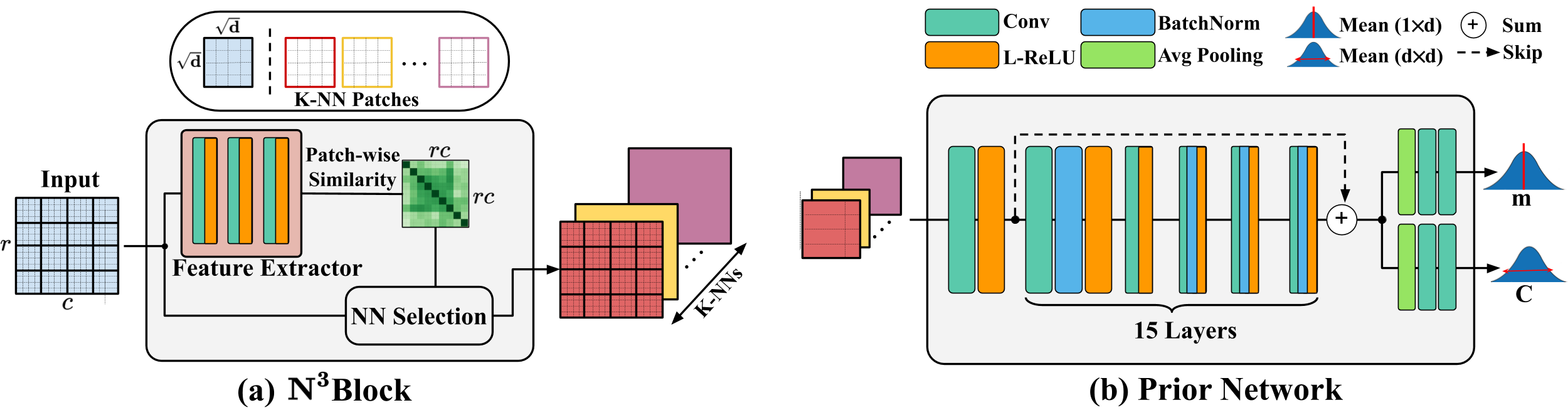}
  \caption{The overall architecture of our proposed network. (a) architecture for $\textup{N}^3$ block, and (b) architecture of prior network}
  \label{fig:overall_arch}
\vspace{-3mm}
\end{figure*}
}

\newcommand\figQual{
\begin{figure*}[t!]

\centering
  \includegraphics[width=\textwidth]{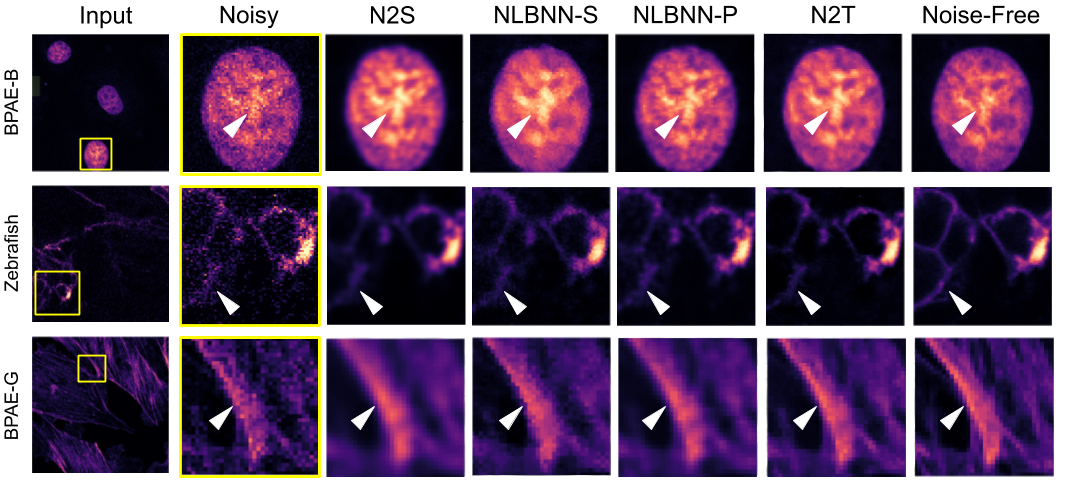}
  \caption{Qualitative results for three confocal microscopy images}
  \label{fig:qual_res}
\vspace{-3mm}
\end{figure*}
}

\newcommand\figKNN{
\begin{figure*}[t!]

\centering
  \includegraphics[width=\textwidth]{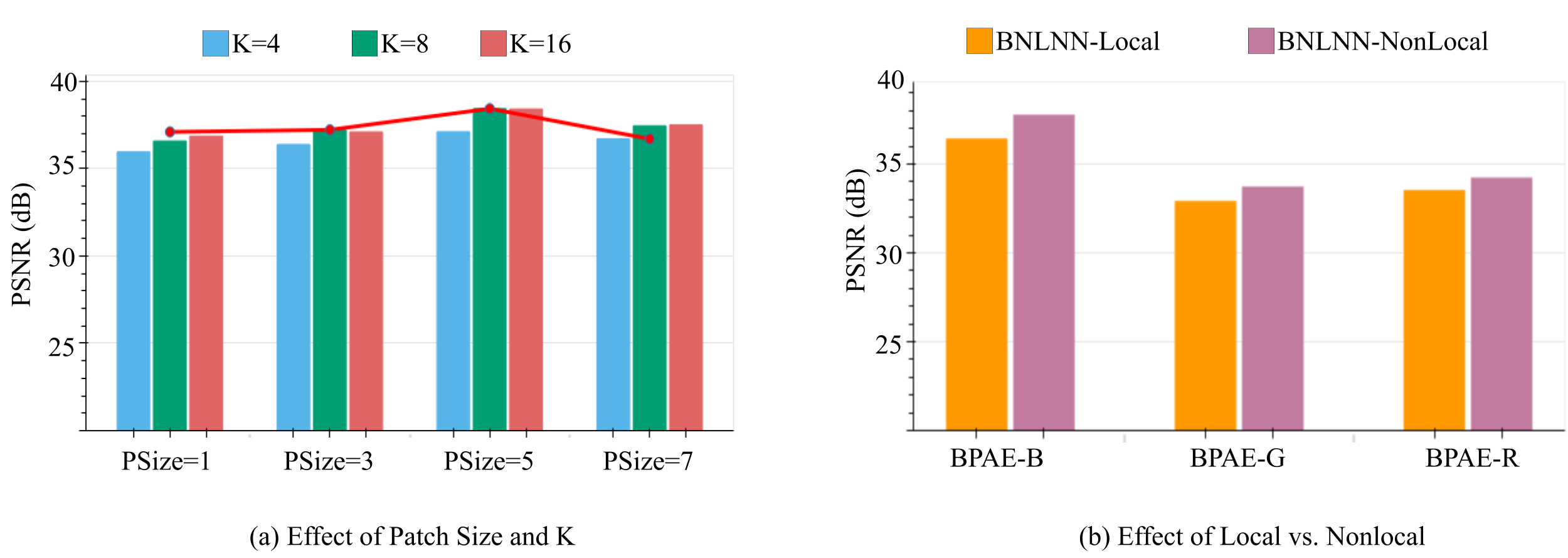}
  \caption{Analysis of different configurations of the proposed model}
  \label{fig:ablation}
\vspace{-3mm}
\end{figure*}
}
\begin{abstract}

Confocal microscopy is essential for histopathologic cell visualization and quantification. Despite its significant role in biology, fluorescence confocal microscopy suffers from the presence of inherent noise during image acquisition. Non-local patch-wise Bayesian mean filtering (NLB) was until recently the state-of-the-art denoising approach. However, classic denoising methods have been outperformed by neural networks in recent years. In this work, we propose to exploit the strengths of NLB in the framework of Bayesian deep learning. We do so by designing a convolutional neural network and training it to learn parameters of a Gaussian model approximating the prior on noise-free patches given their nearest, similar yet non-local, neighbors. We then apply Bayesian reasoning to leverage the prior and information from the noisy patch in the process of approximating the noise-free patch. Specifically, we use the closed-form analytic \textit{maximum a posteriori} (MAP) estimate in the NLB algorithm to obtain the noise-free patch that maximizes the posterior distribution. The performance of our proposed method is evaluated on confocal microscopy images with real noise Poisson-Gaussian noise. Our experiments reveal the superiority of our approach against state-of-the-art unsupervised denoising techniques. 
\par 
\end{abstract}

\section{Introduction}
Confocal fluorescence microscopy (CFM) has become an indispensable tool in cell biology that provides visualization of living cells and tissues, hence forming the basis for the analysis of their morphological and structural characteristics. Nevertheless, the excitation laser power introduces phototoxic side effects on target cells and even organisms~\cite{Jaroslav2017}. Consequently, fluorescence microscopy has to be acquired in low illumination setting, which limits the number of collected photons at the detector plane~\cite{Zhang2019}. Consequently, CFM images are mainly dominated by Poisson noise that renders them less reliable and undermines the biological conclusions drawn therefrom~\cite{Jaroslav2017}. To remedy the problem of low signal-to-noise-ratio (SNR) in CFM, the application of noise reduction methods has become an essential pre-processing steps preceding any diagnostics or other  biological analyses~\cite{Hadj2013,Roudot2013}. \par

Until recently, the `medal' for state-of-the-art image denoising was held by non-local patch-based methods~\cite{nlm2005,bm3d2007}, which exploit the repetitiveness of patch patterns in the image. To denoise a single patch, a common approach is to retrieve its similar patches within a confined neighborhood followed by an averaging operation over pixel intensities across all neighbors. A Bayesian interpretation of the non-local patch-based schemes was proposed by Lebrun et al., which is based on the assumption that nearest neighbor patches are \textit{i.i.d} samples from a multivariate Gaussian distribution approximating the prior distribution of noise-free samples~\cite{nlb2013}. Given the prior and the input observation likelihood at hand, a \textit{maximum a posteriori} estimate results in a Wiener filter which is used to infer the denoised patch. \par

The popularity of deep learning has ignited extensive research aimed at leveraging the capabilities of neural networks for discriminative learning. In the context of denoising, however, a challenging shortcoming of existing discriminative approaches is that their training required noise-free and noisy image pairs. Low-noise images may be collected (in lieu of noise-free images) at the expense of longer acquisition times or more advanced hardware.
In the absence of noise-free images, notable progress has been made by relying on the statistical characteristics of noise and the underlying signal, which led to the introduction of self-supervised learning paradigms for denoising that only require single or pairs of noisy images~\cite{n2v2019,n2s2019,whitenner2019}. Recently, further promising improvements have been achieved by leveraging the Bayesian neural network for pixel-wise probabilistic inference of noise-free values~\cite{prakash2019fully,krull2019probabilistic,Samuli2019}. However, all of these methods construct posterior probability distributions at the pixel level and using only the surrounding local context while ignoring all co-variance between the pixels within patches as well as the valuable source of non-local information across the image. 

\noindent \textbf{Summary of Contributions.} To circumvent the limitations mentioned above, we make the following contributions:
1) We propose a patch-based extension of the previous probabilistic self-supervised denoising methods that do not require  ground truth information; 2) instead of relying on the local context for learning priors, we propose to use the information from multiple non-local patches across the image; 3) we generate similar results to supervised methods even though our method does not observe any noise-free images during training; and 4) our method yields superior performance compared to previous pixel-wise self-supervised approaches.

{\let\thefootnote\relax\footnote{{\scriptsize{We use bold capital letters, bold small letters, and regular small letters to denote matrices, vectors, scalars, respectively. Also,  $\mathbf{A}^T$ and $\mathbf{A}^{-1}$ indicate transpose and inverse of matrix $\mathbf{A}$.}}}}

\section{Theoretical Background}
An observed noisy image $\mathbf{Y}$ can be decomposed into a set of (non)overlapping patches of size $\sqrt{d} \times \sqrt{d}$ denoted by $\mathcal{Y} = \{\mathbf {Y}_{i} \in \mathbb{R}^{\sqrt{d} \times \sqrt{d}}\}_{i=1}^{\textup{N}}$. To simplify the following calculations, all patches are further re-arranged into vectors with $d$ elements. In this setting, an arbitrary patch $\mathbf{y}_i \in \mathcal{Y} \in \mathbb{R}^d$ from the set  can be decomposed as: 
\begin{equation}
\label{eq1}
    \mathbf{y}_i = \mathbf{x}_i + \boldsymbol{\eta}_i,     
\end{equation}

\noindent where $\mathbf{x}_i$ and $\boldsymbol{\eta}_i \in \mathbb{R}^{d}$ indicate the underlying noise-free patch and additive noise component, respectively. Patch-based image denoising refers to the task of inspecting the noisy patches $\mathbf{y}_i$ to infer its corresponding noise-free patch $\mathbf{x}_i$. Let $\mathcal{S}_i = \{\mathbf {y}_{t}\}_{t=1}^{\textup{k}}$  denote $k$ nearest neighbor patches based on the euclidean similarity. To find the estimate of the denoised patch $\mathbf{\tilde{x}}_i$ in Bayesian framework, we need to induce the posterior distribution over all possible noise-free patches conditioned on the set of nearest neighbors $\mathcal{S}_i$ as well as the observed noisy patch $\mathbf{y}_i$. Then, the optimum noise-free estimate is the one which maximizes the posterior distribution, i.e.:

\begin{equation}
\label{argmax}
    \mathbf{\tilde{x}}_i = \mathop{\arg \max}_{\mathbf{x}_i} \hspace{0.1em} p(\mathbf{x}_i|\mathbf{y}_i, \mathcal{S}^i).
\end{equation}
In the following, we present mathematical derivations for a single patch and omit the subscript $i$ for improved readability. According to the Bayes' rule, Eq.~\ref{argmax} can be transformed to: 
\begin{equation}
    \mathbf{\tilde{x}} \propto \mathop{\arg \max}_{\mathbf{x}} p(\mathbf{y}|\mathbf{x}) 
    p(\mathbf{x}|,\mathcal{S}).
\end{equation}

\noindent The term $p(\mathbf{y}|\mathbf{x})$ is the observation likelihood and $p(\mathbf{x}|\mathcal{S})$ captures the prior knowledge we have about a noise-free patch considering its non-local nearest neighbors. We approximate the prior using a multivariate Gaussian distribution $\mathcal{N}(\mathbf{m}_x,\mathbf{C}_x)$ parametrized by $\mathbf{m}_x \in \mathbb{R}^{d}$ and $\mathbf{C}_x \in \mathbb{R}^{d \times d}$.  The maximum likelihood estimates of the prior parameters, over the nearest neighbor patches $\mathbf{y}_t$, are given by:
\begin{equation}
\label{mle}
    \overline{\mathbf{m}}_x = \frac{1}{K} \sum_{t=1}^{k}\mathbf{y}_t
    \quad \textup{and} \quad
    \overline{\mathbf{C}}_x=\frac{1}{k-1} \sum_{t=1}^{k} (\mathbf{y}_t - \overline{\mathbf{m}}_x)(\mathbf{y}_t - \overline{\mathbf{m}}_x)^T.
\end{equation}

\noindent For the generic form of Poisson-Gaussian noise where the observed noisy patch contains both signal-independent and signal-dependent corruption, the noise model is approximated by a heteroscedastic Gaussian model $\mathcal{N}(0, \boldsymbol{\beta}^{2} \mathbf{x} + \sigma^2)$ whose variance is a function of the true noise-free measurement $\mathbf{x}$. Symbols $\boldsymbol{\beta} \in \mathbb{R}^d$ and $\sigma \in \mathbb{R}$ refer to the gain of the signal-dependent and standard deviation of the signal-independent noise components, respectively. In this setting, the observation likelihood follows a Gaussian distribution $\mathcal{N}(\mathbf{m}_y,\mathbf{C}_y)$ and its parameters are estimated as: 
\begin{equation}
\label{lparams}
    \overline{\mathbf{m}}_{x} = \overline{\mathbf{m}}_{y} \hspace{0.5em} \textup{and} \hspace{0.5em}
    \overline{\mathbf{C}}_{y} = \overline{\mathbf{C}}_{x} + \boldsymbol{\beta}^2 \hspace{0.2em} \overline{\textbf{m}}_x \mathbf{I} + \sigma^2\textbf{I},
\end{equation}
where $\mathbf{I}$ denotes a diagonal identity matrix. Following ~\cite{Samuli2019}, we replace the ground truth noise-free patch $\mathbf{x}$ with the predicted prior mean $\overline{\mathbf{m}}_x$ in Eq.~\ref{lparams}. With the prior and observation likelihood at hand, a closed-from MAP estimate can be used to infer the denoised patch maximizing the posterior distribution: 
\begin{equation}
\label{map}
    \mathbf{\tilde{x}} = \overline{\mathbf{m}}_x + 
    \overline{\mathbf{C}}_x  {\overline{\mathbf{C}}_y}^{-1}  (\mathbf{y} - \overline{\mathbf{m}}_x).
\end{equation}

\figArch

\section{Patch-based Non-local Bayesian Networks}

We propose a patch based non-local Bayesian denoising in the scope of deep neural networks. Original non-local Bayesian denoising approach~\cite{nlb2013} uses maximum likelihood (ML) estimates in Eq.~\ref{mle} to approximate the parameters of the assumed multivariate Gaussian prior, however, ML estimates are prone to yielding inaccurate outcomes as they directly manipulate the noisy patch raw intensities in $\mathcal{S}$. To remedy this, we adopt a Bayesian neural network~\cite{gal2017} as a learning based estimator which approximates prior parameters in a more robust way. In particular, we design a prior network which receives the patches in $S$ and regresses the estimates $\overline{\textbf{m}}_x$ and $\overline{\textbf{C}}_x$. With prior and observation likelihood at hand, one can use the MAP estimates to find the optimal approximation of of the denoised patch.

\subsection{Training}

\noindent \textbf{Non-local Similar Patch Search.} To form the similar patch set $\mathcal{S}$ for a reference patch $\mathbf{y}$, we tailor the nearest neighbor network ($\textup{N}^3$Net) proposed by Plötz~\cite{n3net2018}. In $\textup{N}^3$Net, the nearest neighbor selection rule is interpreted as a $k$-way categorical classification over the candidate patches based on their euclidean similarities derive in a learnable feature space. Therefore, to retrieve $k$ similar patches for $\mathbf{y}$, $\textup{N}^3$Net carries out $k$ successive samplings from categorical distribution while discarding the patches already picked for the subsequent rounds. Our proposed  framework differs from $\textup{N}^3$Net~\cite{n3net2018} in three aspects: 1) instead of interleaving multiple $\textup{N}^3$ blocks across intermediate layers, we employ only an individual $\textup{N}^3$ as initial component of our prior network, 2) in contrast to $\textup{N}^3$Net which finds similar patches within the intermediate feature maps, we utilize $\textup{N}^3$ block directly in the spatial domain, and 3) $\textup{N}^3$Net considers categorical logits for all patches as weights and gives a weighted average of all patches as the aggregated nearest neighbor, however we discard the aggregation phase and explicitly pick the most likely patch as the $k$-th nearest neighbor. As depicted in Fig~\ref{fig:overall_arch}-a, the non-local similar patches are concatenated to across the channel dimension. \par

\noindent \textbf{Prior Network.} Equipped with a set of observed patches $\mathcal{D}=\{\mathbf{Y}\}$, their decomposed patches $\mathcal{Y}$, and non-local similar patches $\mathcal{S}$ for every entry in $\mathcal{Y}$, we train a CNN-based prior network, denoted by $f_{prior}(\mathcal{S}; \theta_{prior})$ and parametrized by $\theta_{prior}$, to learn the mapping from non-local similar patches (excluding the observed noisy patch itself) to the Gaussian prior mean and covariance. Mathematically, 
\begin{equation}
    \overline{\mathbf{m}}_x, \overline{\mathbf{C}}_x =  f_{prior}(\mathcal{S}; \theta_{prior}).
\end{equation}
Our proposed prior network which employs a simple neural architecture, possess two tails at the end of its architecture; either providing the desired estimates $\overline{\mathbf{m}}_x$ and $\overline{\mathbf{C}}_x$ estimates. Since the the covariance estimate must strictly fulfill the statistical characteristic of being symmetric and semi-definite, we adopt the notion of Cholesky decomposition and parametrize covariance matrix as the product of its lower-triangular decomposition matrices, i.e. $\overline{\mathbf{C}}_x = \mathbf{L}_{x}\mathbf{L}_{x}^{T}$. In our implementation we ensure that $\mathbf{L}_{x}$ contains positive-valued diagonal entries. 
During the training, prior network weights are optimized by validating the observation likelihood $\mathcal{N}(\mathbf{m}_y,\mathbf{C}_y)$ on the observed patch $\mathbf{y}$. In other words, we use Eq.~\ref{lparams} to yield estimate $\mathbf{m}_y,\mathbf{C}_y$ and minimize the negative log of observation likelihood over the observed noisy patch to guide the learning in favor of predicting accurate estimates $\overline{\mathbf{m}}_x$ and $\overline{\mathbf{C}}_x$, i.e.:
\begin{equation}
    \mathcal{L}(\mathbf{y}, \overline{\mathbf{m}}_y, \overline{\mathbf{C}}_y) =  \frac{1}{2}(\mathbf{y} - \overline{\mathbf{m}}_y)^{T} \hspace{0.15em} {\overline{\mathbf{C}}_y}^{-1}  (\mathbf{y} - \overline{\mathbf{m}}_y) + \frac{1}{2} \textup{log} \hspace{0.25em} |\overline{\mathbf{C}}_y|
\end{equation}
where $|\cdot|$ indicates the determent to hinder the covariance values become large. \par

\noindent \textbf{Noise Level Estimation.} In Eq.~\ref{lparams}, we assume that the noise parameter $\sigma$ and $\boldsymbol{\beta}$ are knwon apriori. However, it is likely to lack this knowledge in real-world denoising situations. To address this, an alternative way is to specify the noise parameter estimation as a part of the optimization procedure and design a network to predict them during the training and inference~\cite{Samuli2019,gal2017}. Specifically, we adopt a CNN to approximate a function $f_{noise}(\mathbf{y}_i;\theta_{noise})$ that regresses the noise parameters estimates $\overline{{\sigma}}$ and $\overline{\boldsymbol{\beta}}$. In this work, we assume that the $\sigma$ is fixed across the entire image while $\mathbf{\beta}$ varies across different patches. Following~\cite{Samuli2019}, we add a small regularization $-0.1(\sigma + \boldsymbol{\beta})$ to the loss in favor of explaining the noise as corruption and not the uncertainty about the noise-free measurement.  \par

\figQual

\subsection{Inference.} After the prior network is learned, we employ it in a Bayesian inference framing to yield the noise-free estimate of the observed noisy patch. Particularly, we firstly use the trained $\textup{N}^3$ block to collect the non-local similar patches which are subsequently piped in to the prior network to predict prior estimates. Secondly, we derive the observation likelihood from Eq.~\ref{lparams}. As stated earlier, we eventually approximate the noise-free patch $\tilde{\mathbf{x}}$ using the MAP estimate in Eq.~\ref{map}. An interesting interpretation of Eq.~\ref{map} is that the MAP estimates primarily expects $\tilde{\mathbf{x}}$ to equal the $\overline{\mathbf{m}}_x$, and adapts the final estimate by taking into account the influence of the observation likelihood.

\noindent \textbf{Dense Patch Denoising.} Patch-based denoising techniques potentially leave block artifacts in the resultant image -- no exception for our proposed method. To overcome this, we adopt a dense-strided patch denoising scheme which partitions the image into densely overlapped $\sqrt{d}\times \sqrt{d}$ patches and noise-free estimates for all of them are collected. Afterward, the denoised patches are combined to construct the full size denoised image by averaging the overlapped regions between denoise patches.\par

\subsection{Network Architecture.} For the prior network, we employ the DnCNN~\cite{dncnn2017} architecture with 17 convolution layers and 64 features interleaved with LeakyReLU and batch normalization. As depicted in Fig.~\ref{fig:overall_arch}-b, a skip connection sums the output of the first convolution with the output of the layer before the outputs. As mentioned earlier, two tails receive the extracted features and perform an average pooling with kernel size $\sqrt{d} \times \sqrt{d}$ and stride $\sqrt{d}$ followed by a $1\times 1$ convolution to produce the desired outputs of size $\overline{\mathbf{m}}_x \in \mathbb{R}^{d}$ and $\overline{\mathbf{C}}_x \in \mathbb{R}^{\frac{d(d+1)}{2}}$. When noise parameters needs to be estimated, we use a similar backbone to prior network with 5 convolution layer to build the noise estimator network. Similar to the prior network, noise estimator network provides two outputs; $\overline{\sigma}$ for the entire image and $\overline{\beta}$ for each patch. In the $\textup{N}^3$, we use three convolutions with kernel size $3 \times 3$ and 64 features to learn the embedding features for euclidean similarities.

\tblResults

\section{Results and Discussion}
In this section, we present a detailed performance evaluation of our method and comparison against a number of unsupervised, self-supervised and supervised methods for denoising confocal microscopy images corrupted with real noise. \par
\noindent \textbf{Implementation Detail.} As all networks are fully convolutional, we train them all on $90 \times 90$ randomly cropped regions in each epoch. For patch-based methods, the patch size is set to $5 \times 5$ with $k=8$ nearest neighbors. All networks are trained for 100 epochs with a batch size of 4 using Adam optimizer with default parameters. The initial learning rate is set to $3e^{-4}$ and is halved every 40 epochs. Except for the supervised denoising methods, all networks are trained using only the observed noisy images both as the input and target.\par

\noindent \textbf{Data.} We use confocal images from the fluorescence microscopy dataset released by Zhang et al.~\cite{Zhang2019}, which consists of raw images corrupted by real Poisson-Gaussian noise. There exists two single-channel confocal sets, namely Mice brain and Zebrafish, and one multi-channel confocal set, i.e. BPAE. Each category also consists of 20 field of views (FOV). Different samples in each FOV correspond to noisy images of the same scene with a different noise realizations. Similar to~\cite{krull2019probabilistic}, we randomly choose 4 images from the held-out 19-th FOV for testing with the remaining FOVs used for training. For the multi-channel images, we report the mean scores calculated on Red, Green and Blue channels separately. \par

\figKNN
\noindent \textbf{Comparison Methods.} We provide evaluations for two variants of our proposed method: NLBNN-S and NLBNN-P that performs denoising at the pixel-level and patch-level, respectively. We compare these variants against traditional non-local mean(NLM)~\cite{nlm2005}, BM3D~\cite{bm3d2007}, Noise2Void(N2S)~\cite{n2s2019}, DnCNN(N2T)~\cite{dncnn2017}, Nois2Noise(N2N)~\cite{n2n2018}, and probabilistic Noise2Void (PN2V)~\cite{krull2019probabilistic}. We borrow the scores for NLM, BM3D, N2N from~\cite{Zhang2019} and PN2V from \cite{krull2019probabilistic}.
\par

\noindent \textbf{Qualitative Evaluation.} We present our evaluation on the test images in Fig.~\ref{fig:qual_res}; it is evident that N2S generates over-smoothed denoised images as it does not leverage observed noisy information during the inference phase. On the other hand, NLBNN-S and NLBNN-P are able to recover finer textures, especially in regions with high contrast. We attribute these improvements to the fact that MAP estimator used in NLBNN framework has the flexibility to adaptively combine information from the prior and observation. Between NLBNN variants, we observe that NLBNN-S brings unwanted non-uniform estimates. This is explained by the fact that NLBNN-S performs pixel-wise image manipulation and therefore lacks the implicit prior on generating regularized estimates within a patch. Conversely, NLBNN-P consistently captures the cross-pixel signal dependencies in a patch, leading to smoother results compared to NLBNN-S. Most importantly, Fig.~\ref{fig:qual_res} reveals the strength of NLBNN-P to produce denoised images close to the ones obtained by DnCNN~\cite{dncnn} even without access to any ground truth data during the training. \par
\noindent \textbf{Quantitative Evaluation.}
Table~\ref{tab:results} summarizes the performance of our proposed framework against a wider range of state-of-the-art-methods using peak-signal-to-noise-ratio (PSNR) and structural similarity (SSIM) across different confocal subsets. In this table, we observe that BNLCNN-P outperform N2S~\cite{n2s2019}, BM3D~\cite{bm3d2007}, and NLM~\cite{nlm2005} in terms of PSNR over BPAE, ZebraFish and MICE by an average of 0.48dB, 0.84dB, and  1.5dB, respectively. However, it still lags behind the DnCNN~\cite{dncnn} and N2N~\cite{n2n2018} which leverage stronger supervision during the training. From Table~\ref{tab:results}, we also notice that classic BM3D, a classic patch-based non-local methods, provides competitive performance against strong deep learning based methods and even outperforms N2S on the Mice Brain set. Finally, we highlight that BNLCNN-P is able to provide superior numerical results against the DnCNN in the ZebraFish set.
\par

\noindent \textbf{Effect of Patch Size and KNN.} We examine the effect of patch size and the number $k$ of non-local similar patches on the denoising performance, over the BPAE sets, while fixing all other parameters. The colored bars in Fig.~\ref{fig:ablation}-a show that the denoising performance always improves as we increase number of retrieved patches from 4 to 8. Further increasing $k$ to 16, however, does not consistently improve performance (it does for patch size 1 but not 3, 5, or 7, wherein PSNR remain almost unchanged). This is not surprising as every additional retrieved patch is gradually less similar to the reference. Therefore, the added computational complexity of retrieving more patches beyond $k=8$ is unjustifiable. Fixing $k=8$ and comparing performance across various patch sizes (red poly-line), we note that using patch size 5 outperforms size 1, which motivates using patches instead of pixel-wise predictions. However,  performance drops for larger patch size 7 since larger-sized patches are more heterogeneous and thus finding more similar patches become less probable. 

\noindent \textbf{Non-Local vs. Local.} Next, we study the role of non-local patch retrieval vs. the local patch selection strategy with $k$ and patch size fixed to 8 and 5, respectively. Particularly, we compare the denoising performance of our proposed NLBNN-NonLocal framework against its NLBNN-Local variant in which the $k=8$ non-local patches are replaced with non-overlapping adjacent (i.e. local) patches. Examining Fig.~\ref{fig:ablation}-b, we observe that denoising with non-local patches consistently outperforms,
across the three BPAE categories, denoising with local patches. \par

\section{Conclusion}
We proposed an effective patch-based non-local algorithm for image denoising using the Bayesian reasoning scheme that only requires noisy observations during the training. Given an observed noisy patch, our proposed network leverages a prior network to approximate the parameters of a Gaussian prior about the noise-free counterpart by considering only its non-local nearest neighbor patches. 
With the prior distribution and the observation likelihood at hand, we use a closed-form MAP estimator to infer the noise-free estimate that maximizes the posterior distribution. In contrast to preceding probabilistic self-supervised methods that adopt pixel-level Bayesian inference, we empirically pointed out the benefits of patch-based Bayesian inference and merits of self-similarity priors captured via non-local patches. Furthermore, we presented the first attempt to leverage a non-local neural network in an unsupervised image denoising context. Overall, our proposed method both quantitatively and qualitatively outperforms state-of-the-art self-supervised and unsupervised methods. 

\noindent \textbf{Acknowledgments}. Thanks to the NVIDIA Corporation for the donation of Titan X GPUs used in this research and to the Collaborative Health Research Projects (CHRP) for funding.
\bibliographystyle{IEEE}
\bibliography{refs}

\clearpage
\end{document}